\documentclass[a4paper,12pt]{article}
\usepackage{amsmath,amssymb,graphicx}
\newcommand{\beq}{\begin{equation}}

\newcommand{\eeq}{\end{equation}}
\newcommand{\myref}[1]{~{(\ref{#1})}}

\def \be  {\begin{equation}}
\def \ee  {\end{equation}}
\def \ba  {\begin{eqnarray}}
\def \ea  {\end{eqnarray}}
\def \l   {\left}
\def \r   {\right}

\begin{document}
\begin{flushright}

ITEP-TH-18/06 \end{flushright}
\begin{center}
{\huge Monopole Decay \\

 in a Variable External
Field}\\
\vspace{1.5cm}
 {\large A. K. Monin$^{\dagger\ddagger}$\\
 {{A. V. Zayakin}}$^{\dagger\ddagger}$\\}
\vspace{1.3cm}
E-mail: \verb"monin@itep.ru, zayakin@theor.jinr.ru"\\
\vspace{0.5cm} $^{\dagger}$ M.V. Lomonosov Moscow State
University,\\ 119992, Moscow,
 Russia\\ \vspace{0.2cm} $^{\ddagger}$Institute for
Theoretical and Experimental Physics\\ 117259, Moscow, B.
Cheremushkinskaya 25, Russia
\end{center}

\vspace{0.5cm}

\abstract{The rate of monopole decay into a dyon and an electron
in an inhomogeneous external electric field is calculated by
semiclassical methods. Comparison is made to an earlier result
where this quantity was calculated for a constant field.
Experimental and cosmological tests are suggested.}\vspace{3cm}
\newpage

\section{Spontaneous and Induced Decay in External Fields}
Spontaneous non-perturbative processes of particle production in QFT
(Schwinger processes), or vacuum decay processes have been studied
since the historic papers  of Euler--Heisenberg and
Schwinger~\cite{Heisenberg:1935qt,Schwinger:1951nm} on $e^+e^-$
generation in a constant electromagnetic field. Voloshin, Kobzarev,
Okun \cite{Kobzarev:1974cp} were the first to treat false vacuum
decay in a scalar field theory with a stable and a metastable vacuum
states. Later Callan and
Coleman~\cite{Coleman:1977py,Callan:1977pt} gave this
problem a 1-loop treatment, calculating both the exponent and the
preexponential factor of false vacuum decay probability.
\par
Consideration of
induced~\cite{Affleck:1979px,SelivanovVoloshin}
non-perturbative particle creation was a
natural extension of the scope of the problems described
above. The
term ``induced'' denotes the situation in which the initial state is
not vacuum, but rather contains some particle(s). False vacuum decay
in a scalar and spinor field theory, induced by presence of an
external particle acting as a ``catalyst'' or ``nucleation center'',
was treated semiclassically up to 1-loop preexponential
in~\cite{Gorsky:2005yq}.

\par On the other hand, generalization of
Schwinger processes description can be thought of as extending the
class of fields in which the appropriate process takes place. The
original Euler and Heisenberg calculation in QED was performed for
a constant field. For harmonic plane waves calculations had first
been done by Schwinger in the cited paper. One can make
sure~\cite{akhiezer} that
 the same expression is true for adiabatically
varying fields. Narozhny and Nikishov~\cite{Narozhnyi:1970uv}
calculated exactly the effective Lagrangian in an electric field,
dependent on time as $E(t)\sim\frac{1}{\cosh^2(\Omega t)}$. A
semiclassical treatment of a broad class of fields was given
in~\cite{Popov:1973az}. Semiclassical methods were further developed
basing upon WKB approximation~\cite{Popov:2001ak} and the so called
``worldline instanton method''~\cite{Affleck:1981ag,Dunne:2005sx,
Dunne:2006st}. To mention some other exact results, Fried and
Woodard \cite{Fried:2001ur} gave an expression for arbitrary
light-cone coordinate dependent field $E(x_0\pm x_1)$.  For a
comprehensive review of recent developments in Euler
--- Heisenberg effective actions the reader may consult Dunne's
review \cite{Dunne:2004nc}. This extensive list (most part of
which has been left behind in order not to overload the reader) of
Schwinger processes in inhomogeneous external fields is
mostly
related to QED processes. Some authors also dealt Hawking
radiation in a Schwinger-like
manner~\cite{Parikh:1999mf,Angheben:2005rm}. It should be
emphasized that most of the papers on inhomogeneous field vacuum
decay consider spontaneous processes.\par

In the present short paper we suggest combining the both
generalizations of Schwinger processes.   The possibility of
an induced
 monopole decay into a dyon and an electron was first
suggested in~\cite{Gorsky:2001up}. Later
it was
calculated in a constant electric field in~\cite{Monin:2005wz} up
to leading classical exponential factor. This problem deserves
attention {\it per se}, but below we also give reasons for
astrophysicists to be interested in such processes.

The paper is organized as follows. In section~\ref{path} we
describe the general techniques of dealing particle decay in terms
of Euclidean 1-particle path integral. The sub-barrier
trajectories and the leading exponential factor are calculated in
section~\ref{expon}.
  The circumstances under which the process
considered might become significant for observers, are
investigated in section \ref{cosm}. In section~\ref{disc} we
discuss our results.

\section{Quasi-Classical Approximation to Path Integral\label{path}}

We are going to study 't Hooft--Polyakov (non point-like) monopole
and dyon. The masses of these particles $M_m$ and $M_d$ are of order
of the scale $\frac{M_W}{\alpha}$ where $M_W$ is generally the scale
at which spontaneous gauge symmetry breaking takes place (the mass
of W-boson), $\alpha$ coupling constant. At the same time, their
sizes are of the order of magnitude $M_W^{-1}$, thus in weak
coupling limit de Broglie's wavelengths of monopole and dyon are far
smaller then corresponding sizes, therefore, these particles are
essentially classical objects.

On the other hand, 't Hooft--Polyakov monopole does not possess a
well-established local field-theoretical description. That is why
it is reasonable to treat monopole and dyon in terms of 1-particle
theory (quantum mechanics), evaluating (bosonic) Feynman path
integrals semiclassically with restriction on the trajectories of
classical motion $r\gg M_W^{-1}$, where $r$ is a typical
trajectory size.  We are going to study monopole propagator in
imaginary (Euclidean) time. To do that semiclassically, one should
first find closed-loop trajectories in Euclidean time, and then
calculate determinants, corresponding to quantum fluctuations
around them. It is essential to take into account only closed
loops contributions because only paths with finite classical
action on them are relevant.
\begin{figure}[t]
\begin{center}
\includegraphics[height = 5cm, width=9cm]{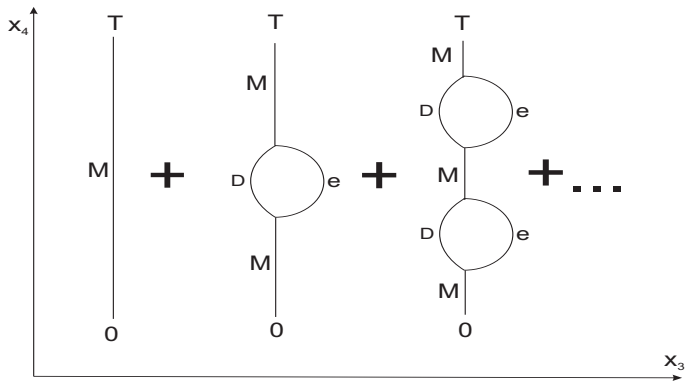}
\caption{ Full 1-loop Green function of a monopole is obtained by
summing over all the insertions of electron-dyon loop into
monopole's Euclidean trajectory.} \label{resumm}
\end{center}
\end{figure}

To find the monopole decay probability in an inhomogeneous
external field one has to calculate corrections to its propagator
in the presence of external electric field. What kind of
trajectories will those corrections correspond to at the classical
level? As monopole is now not forbidden (due to the presence of
the field) to decay into an electron and a dyon,  these are
configurations containing, beside the monopole pieces, dyon and
electron pieces. Therefore, finding the full Green function
of a
monopole in an external field is equivalent to evaluating Green
function with arbitrary number of electron-dyon loop insertions
\beq\notag G(T,0)=G^{(0)}(T,0)+G^{(1)}(T,0)+G^{(2)}(T,0)+\cdots
\eeq The free propagator $G^{(0)}(T,0)$ corresponds to the diagram
without dyon-electron loop insertion, $G^{(1)}(T,0)$ denotes Green
function with one insertion of electron-dyon loop etc., see
Fig.\myref{resumm}. Note that the diagrams of Fig.\ref{resumm} are
not Feynman graphs, but are simply the classical trajectories in
$Ox_4x_3$ plane. However, resummation of all electron-dyon loop
insertions resembles closely the analogous resummation of all
one-particle irreducible vacuum polarization diagram insertions.

\begin{figure}[h]
\begin{center}\label{1loop}
\includegraphics[height = 4cm, width=2cm]{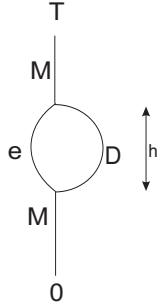}
\caption{One-loop Euclidean configuration (``instanton'' with
winding number $n=1$).}
\end{center}
\end{figure}

\begin{figure}[h]
\begin{center}\label{inst}
\includegraphics[height = 4cm, width=2cm]{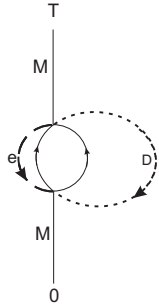}
\caption{One-loop Euclidean configuration with multiply winding
trajectories (``worldline instantons''). Dashed bold arcs
correspond to extra winding paths compared to Fig.\ref{1loop},
arrows indicate winding direction.}
\end{center}
\end{figure}

Taking into account that the electric charge of the magnetic
monopole is zero and using the first quantized approach, the
propagator of the monopole in the Euclidean time can be written as
follows

\beq\notag G^{(0)}(T,0)=\int\mathcal{D}ye^{-M_m\int^T_0
\sqrt{\dot{y}^{2}}d\tau}\sim e^{-M_{m}T}\sqrt{\frac{M}{T^3}}, \eeq
where $y(t)$ is the monopole variable.
 The first correction comes
from Euclidean-time configuration, shown in Fig.~\ref{1loop}.
 \noindent
Note that electron and dyon can go round the loop multiply,
winding over it with some respective winding numbers $m,n$. This
corresponds to the dashed trajectories in Fig. \ref{inst}. Closed
trajectories classified according to their winding number, will be
called, following Dunne and Schubert~\cite{Dunne:2005sx},
``worldline instantons''. They are analogous to Yang--Mills
instanton-antiinstanton pairs.
 \noindent
Propagator correction becomes after resummation over winding
number $m,n$

\beq G^{(1)}_{resummed}(x,z)\sim \sum_{m,n} K_{m,n}
e^{-S^{(m,n)}_{cl}}\label{resummed} \eeq

Here $x=(0,0,0,0),\, z=(T,0,0,0)$ are monopole initial and final
Euclidean coordinates. Equivalent notation $G(x,z)=G(T,0)$ will be
used for such configurations.

To give the reader the feeling of how $S^{(m,n)}_{cl}$ can be
organized, spontaneous vacuum decay case in QED may be brought as an
example. Configurations with greater winding number $n$
would give,
in general, $S^{(n)}_{cl}\sim nS_{cl}$, i.e. $n$ times greater
classical action, and the preexponential factor would depend on $n$
as $K_n=\frac{1}{n^2}K_1$. In the present case, it is reasonable to
expect $S^{(m,n)}_{cl}\sim m S^{(e)}_{cl}+nS^{(d)}_{cl}$.

This resummation is not being performed here, only first term
$K_{1,1}e^{-S_{cl}^{(1,1)}}$ in the sum being left. Below
$K_{1,1}$ will be denoted simply $K$, $S_{1,1}$ as $S_{cl}$. This
resummation would be necessary only in case of $S_{cl}\to 0$,
otherwise the contribution of higher winding number terms is
strongly suppressed. As $S_{cl}\to 0$, both the approximation of
point-like monopole and the semiclassical approximation break
down, so the interest to this resummation is mostly hypothetical.

For the lowest winding numbers $m=1,\,n=1$, the first correction
due to electron-dyon loop is given as \beq G^{(1)}(T,0) =
\int\mathcal{D}ye^{-M_m\int
\sqrt{\dot{y}^{2}}d\tau}\mathcal{D}x\mathcal{D}ze^{-S[x,z,A]},\eeq
where action $S[x(u),z(v),A]$ is the common action for the charged
particles (dyon and electron) in the external field $A_\mu$,
$x(t)$ and $z(t)$ are respectively electron and dyon coordinates,
their worldlines being parametrized by $u,v$
\begin{equation}\notag
S[x,z,A]=m\int\sqrt{\dot{x}^{2}}du+ie\int
A^{ext}(x)\dot{x}du+M_d\int\sqrt{\dot{z}^{2}}dv-ie\int
A^{ext}(z)\dot{z}dv\,\,.
\end{equation}
Monopole coordinate is easily integrated out, as monopole does not
interact with the electric field, \beq \int\mathcal{D}ye^{-M_m\int
\sqrt{\dot{y}^{2}}d\tau}\mathcal{D}x\mathcal{D}ze^{-S[x,z,A]}\sim
\int
e^{-M_{m}(T-h[x,z,A])}\mathcal{D}x\mathcal{D}ze^{-S[x,z,A]},\eeq
here $h$ is the vertical size of the loop (see Fig. \ref{1loop}).

The remaining path integral has a more complicated structure of
modes. It always has zero modes, corresponding to the shifts of
the loop. This will contribute a preexponential factor of
$J[x(t),z(t),A]$, which is the Jacobian, arising due to the
substitution of variables: normalized zero modes coefficients to
the collective coordinates (position of the loop).

The other part of the preexponential factor will come from the
non-zero modes. Among them there will be at least one negative
mode, corresponding to the overall inflation of electron-dyon
loop. Presence of extra negative modes is not evident without a
special investigation~\cite{Monin:2005wz}.  In general, the path
integral yields

\beq\label{kfactor} G^{(1)}(T,0)=\int d^4y G^{(0)}(x,y)
G^{(0)}(y+\Delta y,z) Ke^{-S_{cl}} \eeq where $S_{cl}$ is the
action of dyon and electron in the external field on a classical
path with proper boundary conditions (see Fig.\ref{1loop}). The
integral over $d^4y$ emerges due to integrating over all positions
of the loop, and the two free Green's functions belong to the
purely monopole trajectories. Shift $\Delta y$ accounts for
non-zero size of the loop, and can be neglected as the condition
$T\gg \Delta y$ is imposed.

$K$ contains contributions from the Jacobian and from non-zero
modes \beq\notag K=J[x(t),z(t),A]
\left(\frac{\mathrm{det}F}{\mathrm{det}F_0}\right)^{-\frac{1}{2}}
\eeq Fluctuation operator $F$ can be found in~\cite{Monin:2005wz},
whereas $F_0$ is normalization operator, analogous to operator of
fluctuations around trivial vacuum configuration in false vacuum
decay problems.

We have just given a sketch of structure of the whole expression
for the propagator, as below we are going to be interested mainly
in the leading semiclassical exponent, leaving the preexponential
part for a more sophisticated analysis in future.\par

How can \myref{kfactor} be put into direct correspondence to the
mass shift of the particle? If the full Green's function is
$G(x,z)\approx\frac{1}{(2\pi)^4}\int\frac{e^{ik(x-z)}}{k^2+m^2+\delta
m^2}$, mass shift $\delta m^2$ being imaginary or real, then
expanding in the powers of $\delta m^2$ one gets for the variation
of Green's function

\beq\notag \delta G(x,z)=G^{(1)}(x,z)=-\delta m^2 \int d^4y
G^{(0)}(x,y) G^{(0)}(y+\Delta y,z) \eeq Comparing this with
\myref{kfactor}, one makes sure that

\beq\notag\delta m^2=Ke^{-S_{cl}}\eeq

\section{Exponential factor\label{expon}}
We are going to do path integral over electron and dyon
coordinates by steepest descent approximation, applying the ideas
of world-line instanton method due to Dunne and
Schubert~\cite{Dunne:2005sx,Dunne:2006st} to induced decay
problem. To find the exponential factor of the probability
one should
solve classical equations of motion for dyon and electron in an
external electromagnetic field, find closed configuration of
Euclidean trajectories and minimize the action on them with regard
to trajectory parameters. A single-pulse electric field directed
along  $Ox_3$ axis \beq\notag E_3=E_0\frac{1}{\cosh^2
(\omega
t)}\eeq will be considered.

\begin{figure}
\begin{center}
\includegraphics[height = 5cm, width=9cm]{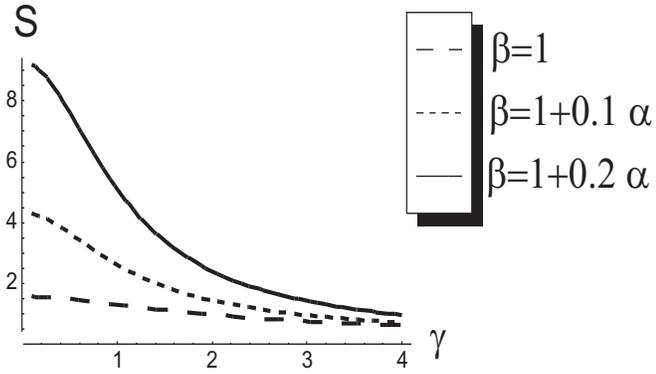}
\caption{Action $S$ of electron-dyon loop in units of
$\frac{m^2}{eE}$ versus Keldysh parameter $\gamma$ for temporally
inhomogeneous field.} \label{stot}
\end{center}
\end{figure}

 Equations of motion will be of the form
\begin{equation}\notag
m\frac{d}{du}\frac{\dot{x}_{\mu}}{\sqrt{\dot{x}^{2}}}=-ie
F_{\mu\nu}(x)\dot{x}_{\nu}
\end{equation}
\begin{equation}\notag
M_{d}\frac{d}{dv}\frac{\dot{z}_{\mu}}{\sqrt{\dot{z}^{2}}}=ie
F_{\mu\nu}(z)\dot{z}_{\nu}
\end{equation}
where $m,M_d$ are electron and dyon masses respectively. Here for
convenience, dimensionless parameters $\gamma=\frac{m\omega}{eE}$,
$\alpha=\frac{m}{M_{d}}$, $\beta=\frac{M_m}{M_d}$ are introduced.
Solution to these equations for the field
$A_{3}(x_{4})=-i\frac{E}{\omega}\tan(\omega x_{4})$ will be for
the electron and for the dyon respectively
\begin{eqnarray}
x_{3}(u)&=&\frac{m}{eE}\frac{1}{\gamma\sqrt{1+\gamma^{2} } }
\textrm{arcsinh}\l(\gamma\cos(2\pi u)\r)-a\notag \\
x_{4}(u)&=&\frac{m}{eE}\frac{1}{\gamma}\arcsin\l(\frac{
\gamma}{\sqrt{1+\gamma^{2}}}\sin(2 \pi u)\r),
u\in\left[-u_{0},u_{0}\right] \nonumber
\\
z_{3}(v)&=&-\frac{m}{eE}\frac{1}{\gamma\sqrt{1+\frac{
\gamma^{2}}{\alpha^{2}}}}
\textrm{arcsinh}\l(\frac{\gamma}{\alpha}\cos(2\pi v)\r)+b\notag \\
z_{4}(v)&=&\frac{m}{eE}\frac{1}{\gamma}\arcsin\l(\frac{
\gamma}{\sqrt{\alpha^{2}+\gamma^{2}}}\sin(2 \pi
v)\r),v\in\left[-v_{0},v_{0}\right], \nonumber
\end{eqnarray}
where $a$, $b$, $u_{0}$, $v_{0}$ are some constants.
\begin{figure}
\begin{center}
\includegraphics[height = 6cm, width=15cm]{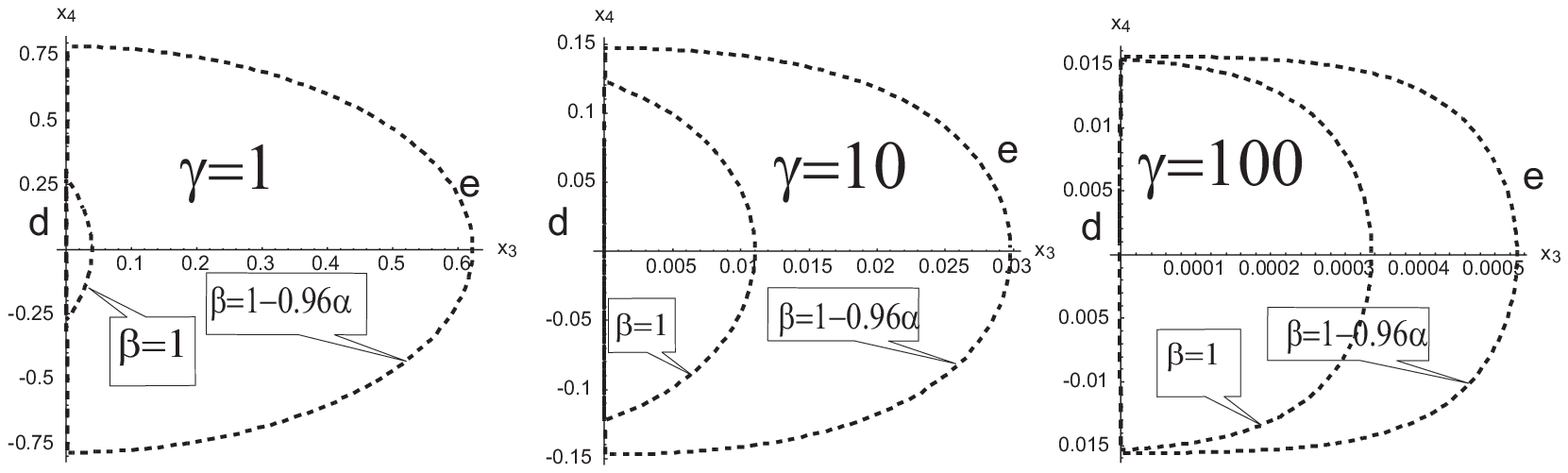}
\caption{Trajectories in the Euclidean plane $(x_3,x_4)$ for
various values of parameters $\beta,\gamma$; $\alpha=5\cdot
10^{-3}$, $\frac{m}{eE}$ taken as length unit.}
\label{trajectories}
\end{center}
\end{figure}
If we calculate the action for these trajectories we will receive a
four-parametric expression, depending on $a$, $b$, $u_{0}$ and
$v_{0}$. We should find the values of these parameters which
minimize the action. One can minimize the action using the standard
procedure and taking into account that resulting loop should be
closed, i.e. $x_4(u_{0})=z_4(v_{0})$. Boundary conditions will in
this case enter the equations as some constraints with Lagrange
multipliers.  Instead of this we can use mechanical analogy
(see~\cite{SelivanovVoloshin}), for which $m$, $M_{d}$ and
$M_{m}$
are forces acting in the vertex, so, the minimum of the action
corresponds to the value of parameters for which we have equilibrium
in the vertex. Condition of the equilibrium is
\begin{eqnarray}
m\cos\alpha_{e}(u_{0})-M_{d}\cos\alpha_{d}(v_{0})&=&0, \nonumber
\\ m\sin\alpha_{e}(u_{0})+M_{d}\sin\alpha_{d}(v_{0})&=&M_{m}.
\label{equilibrium}
\end{eqnarray}
$\alpha_{e}(u_{0})$, $\alpha_{d}(v_{0})$ are the inclination
angles of corresponding trajectory in the vertex. For the electron
we have
\begin{equation}
\tan \alpha_{e}=\sqrt{1+\gamma^{2}}\cot 2\pi u_{0}.
\label{sinalphacotu}
\end{equation}
Using analogous relation for dyon and (\ref{equilibrium}) one can
easily find
\begin{eqnarray}
\tan^{2} 2\pi
u_{0}&=&\left(1+\gamma^2\right)\frac{\left((M_m+m)^2-M_d^2\right)\left(M_d^2-(M_m-m)^2\right)}
{\left(m^{2}+M_{m}^{2}-M_{d}^{2}\right)^{2}},\notag \\
\tan^{2}2\pi
v_{0}&=&\left(1+\frac{\gamma^2}{\alpha^2}\right)\frac{\left((M_m+m)^2-M_d^2\right)
\left(M_d^2-(M_m-m)^2\right)}
{\left(M_{m}^{2}+M_{d}^{2}-m^{2}\right)^{2}}. \notag
\end{eqnarray}
It follows immediately from here that for the boundary conditions
to be properly satisfied, dyon mass should lie within the interval
$M_m-m<M_d<M_m+m$. The action on the extremal trajectory is
\begin{eqnarray}\label{extraction}
S_{cl}(\gamma)=S_{e}(\gamma)+S_{d}(\gamma)-2M_{m}x_{4}(u_{0}),
\end{eqnarray}
where
\begin{eqnarray}
S_{e}(\gamma)&=&\frac{1}{\gamma^{2}}\frac{2\pi m^{2}}{eE}\left(2
u_{0}\sqrt{1+\gamma^2}-\frac{1}{\pi}\arctan\left(\frac{\tan 2\pi
u_{0}}{\sqrt{1+\gamma^2}}\right)-[2u_{0}+\frac{1}{2}]\right)\notag
\\ S_{d}(\gamma)&=&\frac{1}{\gamma^{2}}\frac{2\pi
m^{2}}{eE}\left(2
v_{0}\sqrt{1+\frac{\gamma^2}{\alpha^{2}}}-
\frac{1}{\pi}\arctan\left(\frac{\tan
2\pi v_{0}}{\sqrt{1+\frac{\gamma^2}{\alpha^{2}}}}
\right)-[2v_{0}+\frac{1}{2}]\right), \nonumber
\end{eqnarray}
where square brackets $[x]$ in the last expression denote minimal
integer part of $x$. For $\sin\alpha_{e,d}>0$, or $\tan 2\pi
u_{0}>0$ and $\tan 2\pi v_{0}>0$ (see (\ref{sinalphacotu})) i.e.
$M_{d}^{2}<m^{2}+M_{m}^{2}$ it follows that
\begin{eqnarray}\notag
u_{0}&=&\frac{1}{2\pi}\arctan\sqrt{\left(1+\gamma^2\right)
\frac{\left((M_m+m)^2-M_d^2\right)\left(M_d^2-(M_m-m)^2\right)}
{\left(m^{2}+M_{m}^{2}-M_{d}^{2}\right)^{2}}}<\frac{1}{4},
\nonumber \\
v_{0}&=&\frac{1}{2\pi}\arctan\sqrt{\left(1+\frac{\gamma^2}{\alpha^2}\right)\frac{\left((M_m+m)^2-M_d^2\right)
\left(M_d^2-(M_m-m)^2\right)}
{\left(M_{m}^{2}+M_{d}^{2}-m^{2}\right)^{2}}}<\frac{1}{4},
\nonumber
\end{eqnarray}
so, the action reads as
\begin{eqnarray}
S_{e}(\gamma)&=&\frac{1}{\gamma^{2}}\frac{2\pi m^{2}}{eE}\left(2
u_{0}\sqrt{1+\gamma^2}-\frac{1}{\pi}\arctan\left(\frac{\tan 2\pi
u_{0}}{\sqrt{1+\gamma^2}}\right)\right), \nonumber \\
S_{d}(\gamma)&=&\frac{1}{\gamma^{2}}\frac{2\pi m^{2}}{eE}\left(2
v_{0}\sqrt{1+\frac{\gamma^2}{\alpha^{2}}}-\frac{1}{\pi}\arctan\left(\frac{\tan
2\pi v_{0}}{\sqrt{1+\frac{\gamma^2}{\alpha^{2}}}}\right)\right).
\end{eqnarray}
For the case $M_{d}^{2}>m^{2}+M_{m}^{2}$ it follows that
\begin{equation}
u_{0}=\frac{1}{2}-\frac{1}{2\pi}\arctan\sqrt{\left(1+\gamma^2\right)
\frac{\left((M_m+m)^2-M_d^2\right)\left(M_d^2-(M_m-m)^2\right)}
{\left(m^{2}+M_{m}^{2}-M_{d}^{2}\right)^{2}}}>\frac{1}{4}.
\nonumber
\end{equation}
Corresponding value of the action is
\begin{equation}\label{sresult}
S_{e}(\gamma)=\frac{1}{\gamma^{2}}\frac{2\pi m^{2}}{eE}\left(2
u_{0}\sqrt{1+\gamma^2}-\frac{1}{\pi}\arctan\left(\frac{\tan 2\pi
u_{0}}{\sqrt{1+\gamma^2}}\right)-1\right).
\end{equation}
A family of extremal trajectories for several values of $\gamma$
is shown in Fig. \ref{trajectories}. One can see the overall plot
of $S(\gamma)$ versus Keldysh parameter $\gamma$ in Fig.
\ref{stot}.

\begin{figure}
\begin{center}
\includegraphics[height = 5cm, width=9cm]{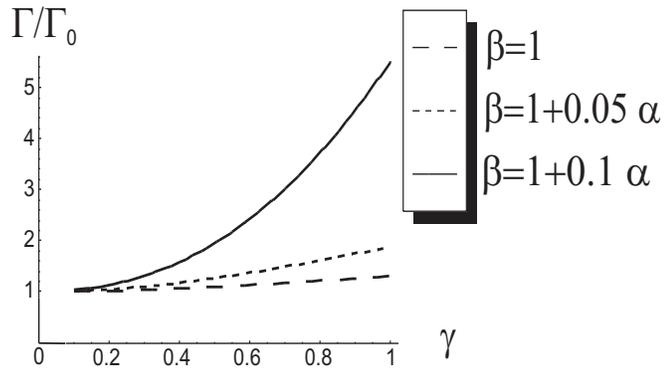}
\caption{$\frac{\Gamma}{\Gamma_0}$ dependence on Keldysh parameter
$\gamma$ for  temporally inhomogeneous field.} \label{w}
\end{center}
\end{figure}

The relation $\frac{\Gamma}{\Gamma_0}$, where $\Gamma_0$ is
monopole mass imaginary part (width) in a constant electric field
($\gamma=0$) can be plotted and is shown in Fig.\ref{w}. One can
easily see the increase of the particle width with the increase of
$\gamma$. This increase becomes stronger when the difference
between $M_d$ and $M_d$ grows.

Action $S$ tends to zero as $\gamma\to\infty$, therefore, at some
point, resummation of higher winding number
contributions~\myref{resumm} will be necessary. At sufficiently
low $S$ the whole process should rather be dealt within common
perturbation theory, because semiclassical approximation to the
non-perturbative mass shift becomes invalid here.

As it was mentioned in section \ref{path} one has a criterion for
applicability of the first quantized approach to dyon and monopole.
Namely, the loop should be large enough for dyon to be treated as a
point-like object. That is, characteristic loop size, which is about
the electronic loop ``radius''
$$r_e=\left\{\begin{array}{l}\frac{1}{\gamma}\frac{m}{eE},\quad
\gamma>>1\\ \frac{m}{eE},\quad \gamma<<1
\end{array}\right.,$$
\noindent should satisfy the condition $$\notag r_e\gg M_W^{-1},
$$ that is,

$$\begin{array}{l}\omega\ll M_W\\
\frac{E_{cr}}{E}>>\frac{m}{M_W}\end{array}$$  Here
$E_{cr}=\frac{m^2}{e}$ is the critical field value for electron.
For imaginable processes in cosmology, field switch-on rate is
obviously less then the enormous 100 GeV $\sim 10^{26}$ Hz, thus
in fact the first criterion is always satisfied. The second
criterion means that $E<<10^{8} E_{cr}$, which is valid for most
of magnetic stars and Reissner-Nordstrom black holes (see section
\ref{cosm}).

However, from general arguments it becomes clear~\cite{Dunne:2005sx}
that semiclassical approximation is wrong when $S\ll 1$. Therefore,
this general limit could give us a more strict criterion of
applicability of the formula~(\ref{extraction}) than monopole size
considerations.

\subsection{Preexponential:  Negative modes}
Here we limit ourselves to qualitative considerations. The
presence of the zero mode was already discussed and used in the
resummation above.
 We might state, due to the fact that for
$\gamma\to 0$ one has a situation, totally analogous to that
described in~\cite{Monin:2005wz}. At sufficiently small $\gamma$ if
$M_d^2<M_m^2+m^2$ one always has  one negative dilatational mode,
corresponding to overall inflation of the loop. It provides a
possibility for monopole mass to acquire an imaginary part. If
$M_d^2>M_m^2+m^2$, another negative mode comes into existence (see
\cite{Monin:2005wz}), thus making loop contribution to the monopole
mass real and describing mass renormalization of a stable monopole.
In future, we are going to elaborate a numerical criterion for
$\gamma$, below which the above statements are true.

\subsection{Spatially inhomogeneous field}
Dunne and Schubert have
shown~\cite{Dunne:2005sx} that in a spatially inhomogeneous
field
particle decay probability can be obtained by an analytic
continuation of the result for the field with the same temporal
inhomogeneity  \beq\gamma\to i\gamma.\eeq See also~\cite{Dunne98,
Fry03} for a more detailed discussion of this analytic
continuation procedure. The general fact on spontaneous (induced)
processes in spatially inhomogeneous fields is that the imaginary
part of the effective action (particle width) decreases with the
increase of $\gamma$, becoming zero at some point. Physically one
can interpret this fact easily: the field characteristic size
becoming too narrow, so that no pair of particles can travel a
path long enough to gain energy necessary for leaving the barrier.
The explicit formula for the action reads
\begin{eqnarray}
S_{cl}(\gamma)=S_{e}(i\gamma)+S_{d}(i\gamma)-2M_{m}x_{4}(u_{0},i\gamma),
\end{eqnarray}
where
\begin{eqnarray}
S_{e}(i\gamma)&=&\frac{1}{\gamma^{2}}\frac{2\pi m^{2}}{eE}\left(-2
u_{0}\sqrt{1-\gamma^2}+\frac{1}{\pi}\arctan\left(\frac{\tan 2\pi
u_{0}}{\sqrt{1-\gamma^2}}\right)\right)\notag
\\ S_{d}(i\gamma)&=&\frac{1}{\gamma^{2}}\frac{2\pi
m^{2}}{eE}\left(-2v_{0}\sqrt{1-\frac{\gamma^2}{\alpha^{2}}}+
\frac{1}{\pi}\arctan\left(\frac{\tan 2\pi
v_{0}}{\sqrt{1-\frac{\gamma^2}{\alpha^{2}}}} \right)\right),
\nonumber
\end{eqnarray}

Following the simple analytic continuation rule, we show how the
exponential factor behaves for field \beq E\sim
\frac{1}{\cosh^2(\omega x_3)}.\eeq

\begin{figure}[ht]
\begin{center}
\includegraphics[height = 5cm, width=9cm]{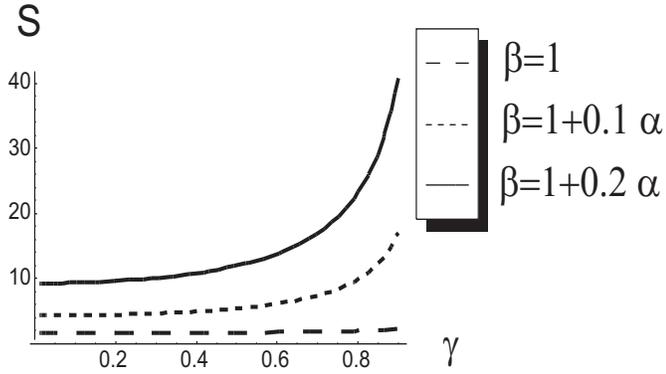}
\caption{Action $S$ of electron-dyon loop in units of
$\frac{m^2}{eE}$ versus Keldysh parameter $\gamma$ for spatially
inhomogeneous field. } \label{stotspace}
\end{center}
\end{figure}

In Fig.~\ref{stotspace} one can see that the process of monopole
decay becomes infinitely suppressed as $\gamma\to 1$. At higher
values of inhomogeneity parameter $\gamma>1$ decay is forbidden.
This corresponds fully to the earlier results
of~\cite{Dunne:2005sx} for spontaneous pair creation.

Below in Fig.\ref{ws} we plot the ratio of monopole width in a
spatially inhomogeneous  field with parameter $\gamma$ to its
width in the constant field case.

\begin{figure}
\begin{center}
\includegraphics[height = 5cm, width=9cm]{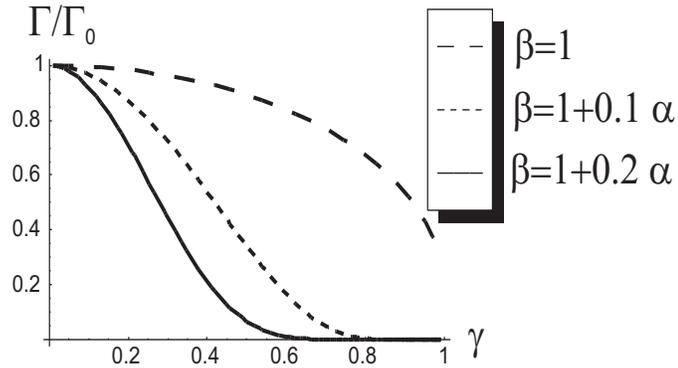}
\caption{$\frac{\Gamma}{\Gamma_0}$ dependence on Keldysh parameter
$\gamma$ for  spatially inhomogeneous field.} \label{ws}
\end{center}
\end{figure}

\section{Monopoles in Cosmology and High-Energy Physics\label{cosm}}
The issue of magnetic monopoles, present in almost every
4-dimensional non-abelian gauge theory with a scalar Higgs field
~\cite{Polyakov:1974ek,'tHooft:1974qc} has long been an important
problem in high-energy physics. Violation of baryon number
conservation law in presence of a magnetic
monopole~\cite{Rubakov:1981tf}, charge
quantization~\cite{Dirac:1948um} are just a few well-known
features of rich monopole physics.\par One of the problems of
monopole physics is the overestimation of monopole concentration
in the Universe with regard to experimental limit. E.g. it has
been pointed out that our Universe should contain relatively large
concentration of free primordial Polyakov -- 't Hooft monopoles of
$M\sim M_W/\alpha\sim 10^4$ GeV (over 10 orders of magnitude
larger than the upper experimental bound), unless they are bound
in meson-like states or there are alternative mechanisms of
monopole decay. GUT monopoles' ($M\sim 10^{17}$ GeV) generation
just after the spontaneous violation of the gauge symmetry of the
GUT group was long being thought to be a major trouble for the
standard cosmological model, until having been solved in terms of
inflationary cosmology in~\cite{Linde:1981mu}. \par
``Phenomenology'' of monopoles nowadays has set some limits on
their abundance. The observed flux $J$ on Earth is limited from
above as $J < 1.4 \times 10^{-16}\frac{1}{cm^2\,s\,sr}$~\cite{
Ambrosio:2002qq}. The so-called Parker bound coming from
magnetohydrodynamic considerations limits the flux of monopoles by
$J<1\times 10^{-15}\frac{1}{cm^2\,sec\,sr}$. The expected
velocities of the monopoles lie in the range $10^{-4}<v/c<1$,
masses of ``primordial'' GUT monopoles are expected to be of the
order of magnitude $10^{17}$ GeV  or higher, monopoles with masses
$10^7\dots 10^{15}$ GeV could have been generated during the later
phase transitions (after breaking the GUT symmetry). Lowest 't
Hooft monopole mass is reached in BPS limit $M\sim M_W/\alpha$.
The limit on concentration of monopoles depends on their velocity
and flux, and can roughly be estimated $n_M<10^{-26}\cdots
10^{-22}\,cm^{-3}$.
\par Thus the concentration of monopoles in the Universe is a very
important quantity, as we have experimental bounds on it and, on the
other hand, the presence of monopoles may have influenced its
evolution. Therefore, it is very important to know well the possible
channels of monopole decay. In particular, one is interested in
external field induced decay processes due to several reasons.\par
First, one cannot fully eliminate the scenario of monopoles decaying
within stellar matter suggested long ago by Zeldovich and
Khlopov~\cite{Zeldovich:1978wj}. One should pay particular attention
to these processes in extremal Reissner--Nordstrom black holes, as
these objects may create electric fields close to the critical
value. (The upper boundary of Reissner--Nordstrom black hole
electric field is actually limited from above by $2\times 10^{12}$
Gs~\cite{Damour:1976jd}). Here one can use constant field limit
$\gamma=0$, and the value of the typical action\myref{extraction} is
$S\sim 10$, so that monopole decay is not infinitely suppressed by
the factor of $e^{-S}\sim
10^{-4.3}$.

\par Second, at the start of inflation stage when monopoles had
just been born, we may have some non-zero expectation value of
(electromagnetic) field, which might have catalyzed monopole
decay
into something else. \par Third, one could consider decay of
monopoles in the interstellar media, as it has been argued that
magnetic fields of the order of magnitude $10^{-2}$ Gs must exist
in dense molecular regions and IR nebulae~\cite{Heiles:1976bz}.

\par Let us estimate realistic values of Keldysh parameter

\beq\notag \gamma=\frac{10^{-19}\,sec}{\tau}\frac{E_{cr}}{E}\eeq
for possible applications of our technique, where $\tau$ is
characteristic inhomogeneity time, $E$ typical field value, and
$E_{cr}$ critical field value for electron ($4\cdot 10^{13}$ Gs).
The most rapid processes observed in the modern Universe have to
do with pulsars and can have $\tau\sim 10^{-3}\,sec$. On the other
hand, typical magnetic field of a pulsar can reach the order of
$10^{-1}E_{cr}$. Thus $\gamma$ is extremely small and one can use
the stationary approximation from~\cite{Monin:2005wz}.

\par In terrestrial conditions, lasers with
$\tau\sim 10^{-15}\cdots 10^{-16}$, and $E\sim 10^{-3} \cdots
10^{-5}$ could be within the reach of modern experimentalists.
Typical action in this case is $S\sim 10^3$, which still
suppresses monopole decay. If gamma-lasers could produce pulses
short enough, one could hope for diminishing $\tau$ by $2-3$
orders of magnitude and reach values $S\sim 1$. However, such
parameters are out of reach at present time.
Moreover, the semiclassical 
approximation breaks down at such values of $S$, 
as we have mentioned already.  Speculating further,
one can put a little fantasy in it and conjecture that
monopoles
could be produced at some high energy facility, and then directed
into a device, where the corresponding time-dependent field is
created. Then monopole decay into a dyon and an electron could be
observed. However, we understand both theoretical and experimental
difficulties of implementing such a project. All attempts to
detect monopoles at accelerators have failed up to now. E.g.,
production of (Dirac) monopoles at Tevatron has been considered
in~\cite{Kalbfleisch:2003yt}. From the data on the existing
facilities, bounds upon monopole mass and cross-section have been
established: $M > 300$ GeV, $\sigma< 10^{-37}$ cm$^{2}$. This
imposes bounds upon the possible monopole production at, e.g.,
LHC. If its luminosity be $\sim 10^{34}$ $cm^{-2}s^{-1}$, the
upper limit of monopole production would be set as 10$^4$  per
year, if we rely on the upper bound for cross-section. It is
obviously a non-trivial task to detect monopole production itself,
and even more complicate done to realize its decay under the
conditions assumed in this paper.

\section{Discussion \label{disc}}

In this short Note the description of the induced monopole decay
has been generalized to the non-stationary field case. Comparing
to stationary field configuration, one can conclude that, as
expected, monopole decay is enhanced by a temporally-inhomogeneous
field and suppressed by a spatially inhomogeneous field. It has
been shown that, despite being non-perturbatively suppressed, this
process may take place under some exotic conditions, e.g. in
Reissner--Nordstrom black holes and in pulse gamma or x-ray
lasers.

Non-stationarity of field becomes a key factor in the latter case,
allowing the classical action on the Euclidean closed worldline to
become sufficiently low and thus cease to suppress monopole decay.
On the other hand, we have shown that even for the most rapid
processes in cosmology, it is generally possible to use constant
field approximation from~\cite{Monin:2005wz}.

\section{Acknowledgements}

Authors are grateful to A. S. Gorsky for suggesting
this
problem and fruitful discussions. This work is supported in
part
by   05-01-00992, Scientific School grant NSh-2339.2003.2
and by JINR Heisenberg --- Landau project (A.Z.);  RFBR
04-01-00646 Grants and Scientific School grant
NSh-8065.2006.2 (A.M.).


\begin{thebibliography}{10}
\bibitem{Heisenberg:1935qt}
  W.~Heisenberg and H.~Euler,
  Z.\ Phys.\  {\bf 98} (1936) 714.

\bibitem{Schwinger:1951nm}
  J.~S.~Schwinger,
  Phys.\ Rev.\  {\bf 82}, 664 (1951).

\bibitem{Kobzarev:1974cp}
  I.~Y.~Kobzarev, L.~B.~Okun and M.~B.~Voloshin,
  Sov.\ J.\ Nucl.\ Phys.\  {\bf 20}, 644 (1975)
  [Yad.\ Fiz.\  {\bf 20}, 1229 (1974)].

\bibitem{Coleman:1977py}
  S.~R.~Coleman,
  Phys.\ Rev.\ D {\bf 15}, 2929 (1977)
  [Erratum-ibid.\ D {\bf 16}, 1248 (1977)].

\bibitem{Callan:1977pt}
  C.~G.~Callan and S.~R.~Coleman,
  Phys.\ Rev.\ D {\bf 16} (1977) 1762.

\bibitem{Affleck:1979px}
  I.~K.~Affleck and F.~De Luccia,
  Phys.\ Rev.\ D {\bf 20}, 3168 (1979).

\bibitem{SelivanovVoloshin}
K. Selivanov and M. Voloshin, ZHETP Lett,42
(1985) 422.


\bibitem{Gorsky:2005yq}
  A.~Gorsky and M.~B.~Voloshin,
  Phys.\ Rev.\ D {\bf 73}, 025015 (2006)
  [arXiv:hep-th/0511095].

\bibitem{akhiezer} A.~I.~Akhiezer and V~.B.~Berestetskiy,
Quantum Electrodynamics [In Russian]. Moscow, 1959,
GIFML, 656 pp.

\bibitem{Narozhnyi:1970uv}
  N.~B.~Narozhnyi and A.~I.~Nikishov,
  Yad.\ Fiz.\  {\bf 11} (1970) 1072
  [Sov.\ J.\ Nucl.\ Phys.\  {\bf 11} (1970) 596].

\bibitem{Popov:1973az}
  V.~S.~Popov and M.~S.~Marinov,
  Yad.\ Fiz.\  {\bf 16}, 809 (1972).


\bibitem{Popov:2001ak}
  V.~S.~Popov,
  JETP Lett.\  {\bf 74}, 133 (2001)
  [Pisma Zh.\ Eksp.\ Teor.\ Fiz.\  {\bf 74}, 151 (2001)].


\bibitem{Affleck:1981ag}
  I.~K.~Affleck and N.~S.~Manton,
  Nucl.\ Phys.\ B {\bf 194}, 38 (1982).

\bibitem{Dunne:2005sx}
  G.~V.~Dunne and C.~Schubert,
  Phys.\ Rev.\ D {\bf 72}, 105004 (2005)
  [arXiv:hep-th/0507174].

\bibitem{Dunne:2006st}
  G.~V.~Dunne, Q.~h.~Wang, H.~Gies and C.~Schubert,
  Phys.\ Rev.\ D {\bf 73}, 065028 (2006)
  [arXiv:hep-th/0602176].

\bibitem{Fried:2001ur}
  H.~M.~Fried and R.~P.~Woodard,
  Phys.\ Lett.\ B {\bf 524} (2002) 233
  [arXiv:hep-th/0110180].

\bibitem{Dunne:2004nc}
  G.~V.~Dunne,
  arXiv:hep-th/0406216.

\bibitem{Parikh:1999mf}
  M.~K.~Parikh and F.~Wilczek,
  Phys.\ Rev.\ Lett.\  {\bf 85}, 5042 (2000)
  [arXiv:hep-th/9907001].
\bibitem{Angheben:2005rm}
  M.~Angheben, M.~Nadalini, L.~Vanzo and S.~Zerbini,
  JHEP {\bf 0505}, 014 (2005)
  [arXiv:hep-th/0503081].



\bibitem{Gorsky:2001up}
  A.~S.~Gorsky, K.~A.~Saraikin and K.~G.~Selivanov,
  Nucl.\ Phys.\ B {\bf 628}, 270 (2002)
  [arXiv:hep-th/0110178].


\bibitem{Monin:2005wz}
  A.~K.~Monin,
  JHEP {\bf 0510}, 109 (2005)
  [arXiv:hep-th/0509047].


\bibitem{Polyakov:1974ek}
  A.~M.~Polyakov,
  JETP Lett.\  {\bf 20} (1974) 194
  [Pisma Zh.\ Eksp.\ Teor.\ Fiz.\  {\bf 20} (1974) 430].

\bibitem{'tHooft:1974qc}
  G.~'t Hooft,
  Nucl.\ Phys.\ B {\bf 79}, 276 (1974).

\bibitem{Rubakov:1981tf}
  V.~A.~Rubakov,
IYAI-P-0211

\bibitem{Dirac:1948um}
  P.~A.~M.~Dirac,
  Phys.\ Rev.\  {\bf 74} (1948) 817.

\bibitem{Linde:1981mu}
  A.~D.~Linde,
  Phys.\ Lett.\ B {\bf 108}, 389 (1982).

\bibitem{Ambrosio:2002qq}
  M.~Ambrosio {\it et al.}  [MACRO Collaboration],
  Eur.\ Phys.\ J.\ C {\bf 25}, 511 (2002)
  [arXiv:hep-ex/0207020].

\bibitem{Zeldovich:1978wj}
  Y.~B.~Zeldovich and M.~Y.~Khlopov,
  Phys.\ Lett.\ B {\bf 79} (1978) 239.

\bibitem{Damour:1976jd}
  T.~Damour and R.~Ruffini,
  Phys.\ Rev.\ D {\bf 14} (1976) 332.

\bibitem{Heiles:1976bz}
  C.~Heiles,
  Ann.\ Rev.\ Astron.\ Astrophys.\  {\bf 14} (1976) 1.


\bibitem{Dunne98} G. V. Dunne and T. Hall, Phys. Rev. D. {\bf 58},
105022 (1998).

\bibitem{Fry03} M. P. Fry. Phys. Rev. D. {\bf 67} 065017 (2003).


\bibitem{Kalbfleisch:2003yt}
  G.~R.~Kalbfleisch, W.~Luo, K.~A.~Milton, E.~H.~Smith and
M.~G.~Strauss,
  Phys.\ Rev.\ D {\bf 69} (2004) 052002
  [arXiv:hep-ex/0306045].




\end{thebibliography}
\end{document}